\documentclass[aps,prb,twocolumn,a4paper,showpacs]{revtex4}
\usepackage{graphicx}

\begin{document}
\title{Er$_2$Ti$_2$O$_7$: Evidence of Quantum Order by
Disorder in a Frustrated Antiferromagnet}

\author{J.~D.~M.~Champion$^{1,2,3}$, M.~J.~Harris$^2$, P.~C.~W.
Holdsworth$^3$, A.~S.~Wills$^4$, G.~Balakrishnan$^5$,
S.~T.~Bramwell$^{1\ast}$, E.~\v{C}i\v{z}m\'ar$^6$, T.~Fennell$^7$,
J.~S.~Gardner$^{8,9}$, J.~Lago$^1$, D.~F.~McMorrow$^10$,
M.~Orend\'a\v{c}$^6$, A.~Orend\'a\v{c}ov\'a$^6$,
D.~M$^{\mathrm{c}}$K.~Paul$^5$, R.~I.~Smith$^2$,
M.~T.~F.~Telling$^2$ and A. Wildes$^4$.}

\affiliation{ $^1$ University College London, Department of
Chemistry, 20 Gordon Street, London WC1H~0AJ, UK.\\
$^2$ ISIS Facility, Rutherford Appleton Laboratory, Chilton,
Didcot, Oxon, OX11~0QX, UK. \\
$^3$ Laboratoire de Physique, Ecole Normale Sup\'erieure, 46
All\'ee d'Italie, F-69364 Lyon, France. \\
$^4$ Institut Laue-Langevin, 6 rue Jules Horowitz, BP 156 - 38042
Grenoble Cedex 9, France. \\
$^5$ Department of Physics, University of Warwick, Coventry,
CV4~7AL, UK. \\
$^6$ Department of Physics, P. J. \v{S}af\'arik University, 041 54
Ko\v{s}ice, Slovakia. \\
$^7$ The Royal Institution of Great Britain, 21 Albemarle Street,
London W1X~4BS, UK. \\
$^8$ Department of Physics, Brookhaven National Laboratory, Upton,
New York 11973-5000, United States.\\
$^9$ NIST Center for Neutron Research, Gaithersburg, Maryland
20899-8562, United States.\\
$^{10}$ Ris\o\ National Laboratory, P.O. 49, DK-4000 Roskilde,
Denmark.}

\begin{abstract}
Er$_2$Ti$_2$O$_7$ has been suggested to be a realization of the
frustrated $\langle 111 \rangle$ XY pyrochlore lattice
antiferromagnet, for which theory predicts fluctuation-induced
symmetry breaking in a highly degenerate ground state manifold. We
present a theoretical analysis of the classical model compared to
neutron scattering experiments on the real material, both below
and above $T_{\rm N} =$1.173(2)~K. The model correctly predicts
the ordered magnetic structure, suggesting that the real system
has order stabilized by zero-point quantum fluctuations that can
be modelled by classical spin wave theory. However, the model
fails to describe the excitations of the system, which show
unusual features.\\
\end{abstract}
 \maketitle

An important aspect of condensed matter is the separation of
energy scales, such that the minimization of one set of
interactions may result in the frustration of another. A paradigm
is the frustrated antiferromagnet, in which the local magnetic
couplings between ions are frustrated by the crystal symmetry that
the ions adopt. However, a systematic study of the rare earth
pyrochlore titanates R$_2$Ti$_2$O$_7$ has shown that local
antiferromagnetic bond frustration is neither a necessary, nor a
sufficient condition for magnetic
frustration~\cite{PRL1,Ramirez,PRL3,Tb,Gd}. Rather, it arises from
the interplay, in the context of the crystal symmetry, of the
principal terms in the spin Hamiltonian. In the case of
R$_2$Ti$_2$O$_7$, the main terms are single-ion anisotropy,
exchange and dipolar coupling. Depending on the balance of these
factors, one observes spin ice behavior (R = Ho,
Dy)~\cite{PRL1,Ramirez,PRL3}, spin liquid behavior (R =
Tb)~\cite{Tb}, and dipole induced partial order (R =
Gd)~\cite{Gd}.

Such behavior is best classified in terms of the dominant $\langle
111 \rangle$ single-ion anisotropy that arises from the trigonal
crystal electric field (CEF) at the rare earth site. For example,
whereas the Heisenberg antiferromagnet has a spin liquid ground
state~\cite{spinliquid}, the $\langle 111 \rangle$ Ising (dipolar)
ferromagnet has a spin ice ground
state~\cite{PRL1,PRL3,denHertogGingras}. There is thus a clear
motivation to study models based on other simple anisotropies and
their realization in the titanate series. In this Letter we study
one such model - the $\langle 111 \rangle$ XY model
antiferromagnet~\cite{BGR} - and its realization
Er$_2$Ti$_2$O$_7$~\cite{Blote,MMM,Siddharthan,Dickon}.

We consider the Hamiltonian:
\begin{equation}\label{Ham}
H = - J \sum_{\langle i,j\rangle} \vec S_i.\vec S_j - D \sum_i
\left(\vec S_i.\vec d_i\right)^2,
\end{equation}
where the classical spins, $\vec S_i$, populate a face centered
cubic array of corner sharing tetrahedra: the pyrochlore lattice.
The spins are confined to easy XY planes by a local $d_i = \langle
111 \rangle$ anisotropy, $D < 0$, and are coupled
antiferromagnetically by exchange $J < 0$. This model was first
studied in Ref.~\cite{BGR}, where a discrete, but macroscopically
degenerate, set of ground states was identified. At finite
temperature thermal fluctuations were found to select an ordered
state by the mechanism that Villain called ``order by
disorder''~\cite{Villain} and a first order phase transition was
observed in numerical simulations. The propagation vector of the
ordered state was found to be ${\bf k}=0,0,0$ (henceforth
``$k=0$''), but the basis vectors of the magnetic structure were
not determined. We have recently discovered that the ground state
degeneracy is more extensive than suggested in Ref.~\cite{BGR};
that there exists a continuous manifold of $k=0$ ground states and
that there may even be disordered states with continuous internal
degrees of freedom~\cite{Dickon}.

The possible basis states of the $k=0$ manifold were identified by
group theory methods. They transform as linear combinations of the
basis vectors of four irreducible representations (IRs), labelled
$\Gamma_{3,5,7,9}$~\cite{Kovalev}. The XY anisotropy energy is
minimized only by (a) linear combinations of the two basis vectors
$\psi_{1,2}$ that transform as the second order IR $\Gamma_5$, (b)
the discrete set of symmetrically equivalent basis vectors
$\psi_{3-5}$ that transform as the third order IR $\Gamma_7$ (see
Fig.~1). Monte Carlo simulations confirmed that the first order
transition at $T_{\mathrm{N}}/J = 0.125$ selects $k=0$ order. An
analysis of the distribution of bond energies suggested that
immediately below $T_{\mathrm{N}}$ there remains a continuous
degeneracy in the $k=0$ manifold~\cite{Dickon}. As $T \rightarrow
0$, the spins were found to settle gradually into the magnetic
structure defined by $\psi_2$ belonging to the continuously
degenerate IR $\Gamma_5$. (Note that $\psi_2$ is the only
non-coplanar structure among $\psi_{1-5}$.) Both the initial
selection of $k=0$ and the final selection of $\psi_2$ must be
order by disorder processes as the ground state manifold is
macroscopically degenerate.

\begin{figure}
\includegraphics[width=0.5\textwidth]{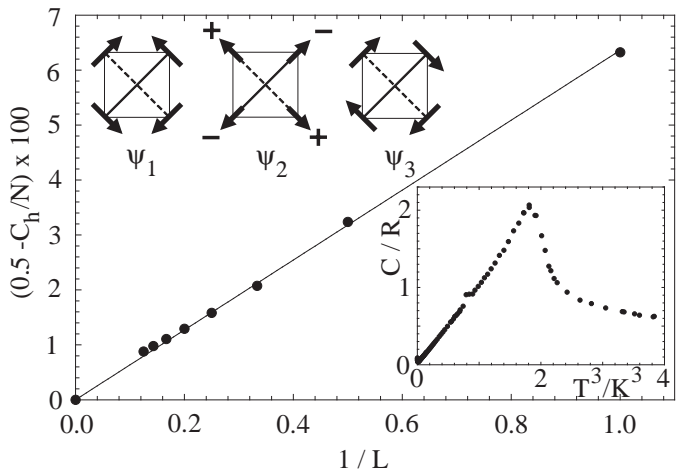}
\caption{\label{fig1}Upper inset: tetrahedral basis projected down
$[001]$, from left to right: $\psi_1,\; \psi_2,\; \psi_3$. $\pm$
denotes how spins tilt out of the plane (they lie parallel to the
opposite triangular face of the tetrahedron). Graph: size
dependence of the simulated specific heat for $N = 16L^3$ spins.
Lower inset: experimental low temperature specific heat {\it vs}
$T^3$ (powder sample). A turn up low at temperature, attributed to
hyperfine effects~\protect\cite{Blote}, is barely visible with
this choice of scale.}
\end{figure}


To understand this ground state selection, we analyze the $D/J
\rightarrow \infty$ model (i.e. spins confined to local $XY$
planes). We expect that for the preferred ground state a spin wave
analysis should expose the presence of zero frequency modes over
an extensive region of the Brillouin zone~\cite{CHS}. This is
indeed the case: we calculate the quadratic Hamiltonian for small
displacements away from a given ground state, which we symmetrize
and diagonalize to find the normal mode spectrum~\cite{Dickon}.
Applying this procedure to the state $\psi_2$ gives eigenvalues
\begin{eqnarray}
\lambda(\vec q)& =& 4J \left(1 \pm \cos\left({\vec q\cdot\vec
a/{2}}\right)\right) \nonumber
\\
\lambda(\vec q)& =& 4J \left(1 \pm \cos\left({\vec q\cdot(\vec c
-\vec b)/{2}}\right)\right),
\end{eqnarray}
where $\vec a$, $\vec b$ and $\vec c$, are the basis vectors of
the primitive rhombohedral unit cell.
Hence, there are branches with $\lambda(\vec q)=0$ over planes in
the Brillouin zone, for which $\vec q\cdot\vec a = 0$ and $\vec
q\cdot(\vec c -\vec b) = 0$. The same procedure when applied to
other selected ground states yields a microscopic number of zero
modes at specific points in the zone. This difference gives the
mechanism for the order by disorder selection of $\psi_2$. In this
approximation the amplitude of the soft modes diverges, giving a
dominant contribution to the entropy~\cite{CHS}.
Evidence of the soft modes exists in the specific heat, as each
contributes less than $\frac{1}{2} k_B$. As there are $O(L^2)$
modes, the quantity $\frac{1}{2} - C_h/Nk_B$ should scale as $1/L$
at low temperature, as confirmed in Fig.~1. We note that the
entropy contribution to the free energy from the soft modes scales
as $N^{\frac{2}{3}}$ and so is not extensive. While this could
mean that the ordering within the $k=0$ manifold occurs at a
temperature-dependent system size that goes to zero in the
thermodynamic limit, no such effect was detected in the system
sizes we have studied.

Disordered states are occasionally formed in the simulations, by
the rotation of columns of spins with infinite length out of an
ordered state. This suggests that, starting from the $\psi_2$
state, one can introduce $O(L^2)$ independent column defects, all
perpendicular to a given plane. Our calculation, giving $O(L^2)$
soft modes is compatible with this description and is analogous to
the case of the Heisenberg kagom\'e antiferromagnet~\cite{CHS}.
For the latter, fluctuations out of a coplanar spin configuration
can be described equivalently in terms of soft propagating modes
and localized zero energy excitations.


\begin{figure}
\includegraphics[width=0.5\textwidth]{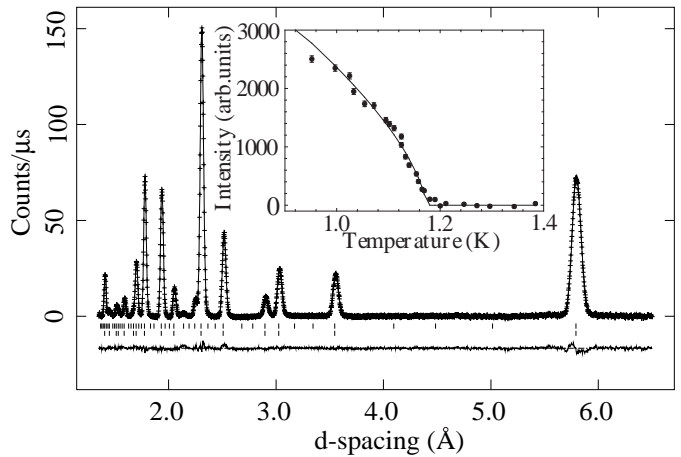}
\caption{\label{fig2} Main picture: powder neutron profile
refinement (POLARIS, 50 mK, $R_{wp} = 1.23~\%$, about a third of
the intensity is magnetic). The lower line is observed minus
calculated intensity. Inset: single crystal Bragg intensity
(PRISMA, (2,2,0) reflection) and fit to a power law (see text).}
\end{figure}


The material Er$_2$Ti$_2$O$_7$, which orders magnetically at $\sim
1.2$ K~\cite{Blote}, has been suggested to approximate the
$\langle 111 \rangle$ XY antiferromagnet~\cite{MMM,Siddharthan}.
To test this we have determined its magnetic structure by powder
neutron diffraction using the POLARIS diffractometer (ISIS). The
magnetic reflections observed below $T_{\mathrm{N}} \approx 1.2$ K
index with a propagation vector of $k=0$. As the transition is
continuous (see below), the system is expected to order under only
one of the non-zero IRs of the Er site representations:
$\Gamma_{3,5,7,9}$, as defined above. Refinement of the magnetic
structure~\cite{Wills} showed that only the two basis vectors
$\psi_1$, $\psi_2$ of $\Gamma_{5}$ were consistent with the
magnetic intensity (see Fig.~1, Fig.~2). Single crystal
diffraction data collected on the instruments E2 (HMI, Berlin) and
PRISMA (ISIS), allowed us to distinguish between the two
structures. The measurements were performed on a ($\sim 8$ mm$^3$)
crystal at temperatures down to 0.13~K. In order to suppress the
formation of multidomains due to the cubic symmetry, a magnetic
field was applied along the [1 $\bar{1}$ 0] direction. Below
$T_{\rm N}$, we found that a field of 0.5~T caused the (2,2,0)
magnetic Bragg peak to increase from 260 counts to 500 counts,
while the other peaks remained approximately unchanged. This
increase in the (2,2,0) by a factor of 1.9 $\pm$ 0.2 is consistent
with the formation of a monodomain of the $\psi_2$ ground state.
We can conclude that the zero field ordering pattern is also
described by $\psi_2$, in agreement with the theory.

The ordered moment at 50~mK (where ordering is essentially
complete) is $3.01 \pm 0.05 ~\mu_B$ per atom.
A CEF analysis ~\cite{CEF}, following Ref.~\cite{Rosenkranz},
predicts a single-ion Kramers doublet ground state with the
wavefunction $-0.5428|-\frac{11}{2}\rangle
-0.2384|-\frac{5}{2}\rangle + 0.5628|\frac{1}{2}\rangle +
0.3876|\frac{7}{2}\rangle - 0.426|\frac{13}{2}\rangle$. This
corresponds to moments of 3.8~$\mu_B$ and 0.12~$\mu_B$
perpendicular and parallel to $\langle 111 \rangle$. In an
antiferromagnet, zero-point quantum fluctuations significantly
reduce the ordered moment from the single-ion value. The observed
moment of $3.01~\mu_B$, thus, agrees well with quasi-classical
ordering of local XY-like moments with a magnitude fixed by the
CEF.

The dominant perturbation to the model Hamiltonian in
Er$_2$Ti$_2$O$_7$ are dipole-dipole
interactions~\cite{Siddharthan} and one might speculate that these
are the true cause for the $\psi_2$ ordering. In the Heisenberg
pyrochlore antiferromagnet, provided the ratio of dipolar energy
to near neighbor exchange is less than a critical value,
$J_{dd}/J_{nn} < 5.7$, the ground state is an energetically
selected $k=0$ state: the $\psi_3$ basis
(Fig.~1)~\cite{Palmer,Palmer2}. As this state is also a ground
state for the $\langle 111 \rangle$ XY model, it is clear that in
the present case it will be stabilized by dipolar interactions. We
expect Er$_2$Ti$_2$O$_7$ to have $J_{dd}/J_{nn} << 5.7$, and thus
$\psi_3$ order, contrary to observation. With a moment of
$3~\mu_B$ the near neighbor dipolar interaction has a magnitude
$-0.32$~K per spin for $\psi_3$ and $+0.06$~K per spin for
$\psi_{1,2}$. It is therefore a substantial fraction of
$T_{\mathrm{N}}$. However, zero point quantum fluctuations also
stabilize the ground state of an antiferromagnet and will favor
the softer $\psi_2$ state~\cite{nnn}.

We are therefore drawn to the conclusion that the most likely
cause of the observed $\psi_2$ order in Er$_2$Ti$_2$O$_7$ is its
stabilization by zero-point quantum fluctuations, the effect of
which is captured by our classical spin wave calculation. However,
the close agreement of experiment and theory does not extend to
the excitations of the system, as we now discuss.

In our model, the density of classical spin wave states,
$g(\lambda)$ is a constant, a result that is incompatible with the
experimental specific heat, which shows a $T^3$ dependence up to
$\sim 1$ K (inset, Fig.~1). Furthermore, in contradiction with
theory~\cite{BGR}, the ordering transition was found to be
continuous within experimental precision. The evolution of the
magnetic Bragg peaks near the critical temperature $T_{\mathrm{N}}
= 1.173(2)$ K obeys a standard power law with the critical
exponent $\beta = 0.33(2)$, characteristic of the 3D-XY model
(inset, Fig.~2)~\cite{precision}. This observation of a
conventional $\beta$ provides a counter example to the
idea~\cite{Reimers} that pyrochlore antiferromagnets represent a
{\it new} universality class, analogous to Kawamura's chiral
universality class for triangular lattice
antiferromagnets~\cite{Kawamura}.


\begin{figure}
\includegraphics[width=0.5\textwidth]{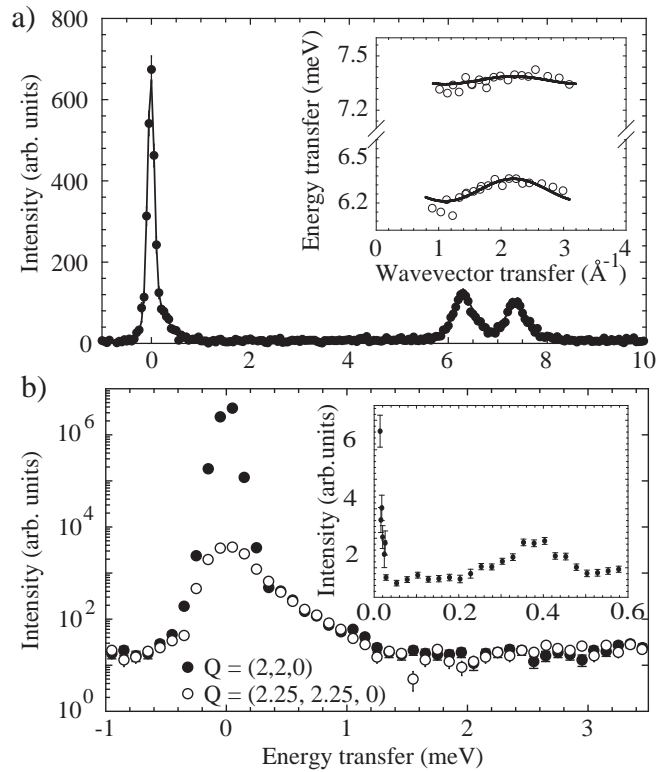} \caption{\label{fig3}a) Powder
inelastic spectrum (PRISMA, 1.8 K, $|{\bf Q}|$ = 1.1 \AA$^{-1}$).
Inset: wave vector dependence of the excitation energies fitted to
a cosine variation. b) Single crystal inelastic scattering (IN14,
100mK, the strong elastic signal is Bragg scattering). Inset:
Powder inelastic scattering (IRIS, 50 mK), integrated over 15
spectra from $|{\bf Q}| = 0.4$ to 1 \AA$^{-1}$. }
\end{figure}


To investigate the excitations further, the dynamics of
Er$_2$Ti$_2$O$_7$ have been investigated by inelastic neutron
scattering. Measurements were carried out on the spectrometers
PRISMA (ISIS), RITA (Ris\o\ National Laboratory) and IN14 (ILL).
For a powder sample, excitons~\cite{exciton} were found at 6.3 and
7.3~meV (Fig.~3a). Their dispersion can be fitted to $E(Q) =
\sqrt{\Delta^2 - 2a^2J(Q)\Delta}$ with $\Delta = 6.38(1)$~meV,
$a\sqrt{J(0)} = 0.35(1)$~meV$^{1/2}$ for the lower level and
$\Delta = 7.39(1)$~meV, $a\sqrt{J(0)} = 0.21(1)$~meV$^{1/2}$ for
the upper level. Here, $J(Q)$ is the Fourier transformed exchange
and $a$ is related to the matrix elements that connect the ground
and excited states~\cite{Grover,comment}. The values of $\Delta$
compare with the CEF predictions of 6.4~meV and
8.8~meV~\cite{CEF}, while $J(0)/k_B$ is implied to be at least
1.5~K. Dispersed excitons are also observed in
Tb$_2$Ti$_2$O$_7$~\cite{Tb}.

In the lower energy scale, data from an image-furnace grown single
crystal ($\sim 250$ mm$^3$) reveal a shoulder of scattering on the
elastic line extending out to $\sim1.5$~meV (Fig.~3b). This
shoulder, which does not appear to obey a strong wave vector
dependence (Fig.~3b), gradually weakens upon heating above $T_{\rm
N}$.
High resolution scans on PRISMA ($T=0.08$~K) and on a powder
sample on the IRIS spectrometer at ISIS (inset, Fig.~3b) resolved
the shoulder into a peak at 0.4~meV, with a width outside the
instrumental resolution.

One possible explanation, that this peak is a very low lying CEF
level, as invoked by Bl\"ote {\it et al.}~\cite{Blote} to explain
the anomalously large Curie-Weiss temperature, $\theta=-22$~K, is
not consistent with the predicted CEF scheme \cite{CEF}. A more
natural explanation is that it is a weakly dispersed optical
magnon mode. Evidence for a gapped magnon spectrum is provided by
the IRIS data (inset, Fig.~3b), which is at the level of
background between 0.03~meV ($\sim 0.3$~K) and 0.2~meV ($\sim 2$
K), indicating an absence of magnons in this range. However, there
is a problem with this interpretation: the $T^3$ specific heat
(inset, Fig.~1) requires a density of ungapped excited states that
increases quadratically with energy up to at least $\sim 1$~K. One
possible way out of this conundrum is to invoke the existence of a
hidden branch of excitations that do not couple directly to
neutrons. It is interesting to note that frustrated, quasi-two
dimensional SrCr$_x$Ga$_{12-x}$O$_{19}$ and
(H$_3$O)Fe$_3$(SO$_4$)$_2$(OH)$_6$ have $T^2$ specific heats
despite the absence of conventional magnons~\cite{SCGO,Wills2}.

A final point of interest is that the other erbium pyrochlores
Er$_2$GaSbO$_7$ and Er$_2$Sn$_2$O$_7$ apparently do not order down
to 50~mK~\cite{Blote, Matsuhira}. It is possible that in this
family there is a kind of quantum critical point where as a
function of the interionic couplings, quasi-classical order gives
way to the quantum fluctuations of the single ion doublets.
Similar effects might also be relevant for
Tb$_2$Ti$_2$O$_7$~\cite{Tb,Mirebeau},
Yb$_2$Ti$_2$O$_7$~\cite{Hodges} and spinels such as
ZnCr$_2$O$_4$~\cite{Zn}.

In conclusion, the simplest theoretical model of
Er$_2$Ti$_2$O$_7$, in which the single-ion moment is fixed by the
CEF, the dipolar interaction is ignored and the exchange is
treated quasi-classically, yields the correct ground state
properties and furnishes strong evidence of quantum order by
disorder. However, it fails to describe the unusual excitation
spectrum of the system and does not encompass the behavior of
Er$_2$GaSb$_2$O$_7$ and Er$_2$Sn$_2$O$_7$. A detailed
understanding of Er$_2$Ti$_2$O$_7$ thus leaves an intriguing
challenge for future research.

It is a pleasure to thank H.-B. Braun, B. Canals, M. J. P.
Gingras, J. A. Hodges, and C. Lacroix  for useful discussions and
R. Down for technical assistance.\\

$^\ast$ Author for correspondence: s.t.bramwell@ucl.ac.uk


\begin{thebibliography}{99}

\bibitem{PRL1}
M.~J. Harris {\it et al.},
\newblock{Phys. Rev. Lett.}
{\bf 79}, 2554 (1997).

\bibitem{Ramirez} A. P. Ramirez {\it et al.},
\newblock{Nature}
{\bf 399}, 333 (1999).

\bibitem{PRL3}
S.~T. Bramwell {\it et al.},
\newblock{Phys. Rev. Lett.}
{\bf 87}, 047205 (2001).

\bibitem{Tb}J. S. Gardner {\it et al.},
\newblock{Phys. Rev. Lett.}
{\bf 82}, 1012 (1999).


\bibitem{Gd} J.~D.~M. Champion {\it et al.}
\newblock{Phys. Rev. B}
{\bf 64}, 140407(R) (2001).

\bibitem{spinliquid} R. Moessner and J.~T. Chalker,
\newblock{Phys. Rev. Lett.}
{\bf 80}, 2929 (1998); B. Canals and C. Lacroix,
\newblock{Phys. Rev. B}
{\bf 61}, 1149 (2000).

\bibitem{denHertogGingras}
B.~C. den Hertog and M.~J.~P. Gingras,
\newblock{Phys. Rev. Lett.}
{\bf 84}, 3430 (2000).

\bibitem{BGR}
S.~T.~Bramwell, M.~J.~P.~Gingras, and J.~N.~Reimers,
\newblock{J. Appl. Phys.},
{\bf 75}, 5523 (1994).

\bibitem{Blote}
H.~W.~J. Bl\"{o}te {\it et al.},
\newblock{Physica}
{\bf 43}, 549 (1969).

\bibitem{MMM}
M.~J. Harris {\it et al.},
\newblock{J. Magn. Magn. Mater.}
{\bf 177}, 757 (1998).

\bibitem{Siddharthan}
R. Siddharthan {\it et al.},
\newblock{Phys. Rev. Lett.}
{\bf 83}, 1854 (1999).

\bibitem{Dickon} J.~D.~M. Champion, Ph.D.
Thesis, Univ. London, (2001).

\bibitem{Villain} J.~Villain, J.
Physique {\bf 41}, 1263 (1980).



\bibitem{Kovalev}
These are a subset of the nine IRs, $\Gamma_{1 \dots 9}$, that
describe all possible $k=0$ spin configurations in the
crystallographic space group $Fd\bar{3}m$. The numbering follows
the scheme defined in O. V. Kovalev, {\it Representations of the
Crystallographic Space Groups} Edition 2 (Gordon and Breach
Science Publishers, Switzerland, 1993).

\bibitem{CHS} J.~T. Chalker, P.~C.~W. Holdsworth, and E.~F.~Shender,
  Phys. Rev. Lett. {\bf 68}, 855,
1992.

\bibitem{Wills}
Using the SARA$h$ program, A.~S. Wills, Physica B {\bf 276-278},
680 (2000), in combination with GSAS, A.C. Larsen and R.B. von
Dreele, General Structure Analysis System (Los Alamos National
Laboratory, 1994).


\bibitem{nnn}
Another possible perturbation to the simple model, further
neighbor exchange interactions, can be shown by symmetry not to
break the degeneracy of the $k=0$ ground state (J.~D.~M. Champion,
unpublished).

\bibitem{Reimers} J.~N. Reimers {\it et al.},
\newblock{Phys. Rev. B} {\bf 45}, 7295 (1992).

\bibitem{Kawamura}
H.~Kawamura,
\newblock{J. Phys. Soc. Jpn.} {\bf 58}, 584
(1989).

\bibitem{Palmer} S.~E. Palmer and J.~T. Chalker,
\newblock{Phys. Rev. B}
{\bf 62}, 488 (2000).

\bibitem{Palmer2} If $J_{dd}/J_{nn} > 5.7$,
a state with $k \ne 0$ is favoured~\cite{Palmer}.

\bibitem{CEF} We thank M.~Rams and J.~A. Hodges for provision
of details of the CEF and for a useful discussion concerning the
single-ion moment.

\bibitem{Rosenkranz}S. Rosenkranz {\it et al.},
\newblock{J. Appl. Phys.}
{\bf 87}, 5914 (2000).



\bibitem{precision}The data suggest a slight systematic increase in $\beta$ as T$_N$ is
approached, with $0.31 < \beta < 0.36$.

\bibitem{exciton} M.~T. Hutchings in {\it Electronic States of
Inorganic Compounds: New Experimental Techniques}, Ed. P. Day,
Nato ASI Series, Vol. 20, (D. Reidel, Dortrecht, 1974).

\bibitem{Grover} B. Grover,
Phys. Rev. {\bf 140}, A1944 (1965).

\bibitem{comment} Modelling of the
dispersion would require a full and very accurate knowledge of the
CEF wavefunctions~\cite{exciton}.

\bibitem{SCGO} A.P. Ramirez, G.P. Espinosa,
and A.S. Cooper,
\newblock{Phys. Rev. Lett.}
{\bf 64}, 2070 (1990).

\bibitem{Wills2} A.~S. Wills {\it et al.}, Europhys. Lett.
{\bf 42}, 325 (1998).

\bibitem{Matsuhira} K. Matsuhira {\it et al.}, J. Phys.
Soc. Japan {\bf 71} 1576 (2002).

\bibitem{Lee} S.-H. Lee {\it et al.},
\newblock{Phys. Rev. B}
{\bf 56}, 8091 (1997).


\bibitem{Mirebeau} I. Mirebeau {\it et al.}, Nature {\bf
420} 54 (2002).

\bibitem{Hodges} J. Hodges {\it et al.}, Phys. Rev.
Lett. {\bf 88} 077204 (2002).

\bibitem{Zn}S.-H. Lee {\it et al.},
\newblock{Phys. Rev. Lett.}
{\bf 84}, 3718 (2000).

\end{thebibliography}
\end{document}